\documentclass[aps,preprint]{revtex4}
\usepackage[english]{babel}
\usepackage{graphicx,dcolumn,bm,amssymb,amsmath,amsfonts,amsthm,latexsym}
\usepackage{epstopdf}
\usepackage{ulem}
\usepackage{float,subfig,slashed,hyperref}
\usepackage[toc]{appendix}
\hypersetup{
  linktocpage  = true,
  colorlinks   = true,
  urlcolor     = red,
  linkcolor    = black,
  citecolor    = blue
}

\begin{document}
\title{Isospin asymmetric matter in a nonlocal chiral quark model}
\author{J.P.~Carlomagno$^a$}
\email{carlomagno@fisica.unlp.edu.ar}
\author{D.~G\'omez~Dumm$^a$}
\author{N.N.~Scoccola$^{b,c}$}
\affiliation{$^{a}$ IFLP, CONICET $-$ Departamento de F\'{\i}sica, Facultad de Ciencias Exactas,
Universidad Nacional de La Plata, C.C. 67, 1900 La Plata, Argentina}
\affiliation{$^{b}$ CONICET, Rivadavia 1917, 1033 Buenos Aires, Argentina}
\affiliation{$^{c}$ Physics Department, Comisi\'{o}n Nacional de Energ\'{\i}a At\'{o}mica,
Avenida del Libertador 8250, 1429 Buenos Aires, Argentina}

\begin{abstract}
We analyze the features of strongly interacting matter in the presence of
nonzero isospin chemical potential $\mu_I$, within a nonlocal two-flavor
Polyakov-Nambu-Jona-Lasinio (PNJL) model. For a system at finite temperature
$T$, we describe the behavior of various thermodynamic quantities and study
the phase diagram in the $\mu_I - T$ plane. In particular, it is found that
for values of $\mu_I$ larger than the pion mass and temperatures lower than
a critical value of about 170~MeV the system lies in an isospin symmetry
broken phase signaled by the presence of a nonzero pion condensate. Our
results for the phase diagram are found to be in better agreement with those
arising from lattice QCD calculations, as compared to the predictions from
other theoretical approaches like the local PNJL model.
\end{abstract}

\maketitle

\section{Introduction}
\label{intro}

The phase diagram of strongly interacting matter at finite temperature and
chemical potential has been extensively studied along the past decades. In
the region of very high temperatures and low densities, it is well known
that Quantum Chromodynamics (QCD) predicts the formation of a quark-gluon
plasma (QGP)~\cite{Fukushima:2010bq}. Under these extreme conditions quark
and gluons are expected to be weakly coupled, and the phase diagram can be
explored by means of first-principle perturbative calculations based on
expansions in powers of the QCD coupling constant. Moreover, lattice QCD
(LQCD) calculations indicate that at vanishing chemical potential the
transition from the hadronic phase to the QGP occurs in the form of a smooth
crossover, at a pseudocritical temperature $T_{\rm pc} \sim 150-170$~MeV. On
the other hand, at sufficiently high densities and low temperatures, one
expects to find a ``color-flavor locked'' phase~\cite{Alford:2007xm}, in
which quarks are bounded into color superconducting states analogous to the
Cooper pairs formed by electrons in an ordinary superconductor. At moderate
densities, however, the situation is much more uncertain. The main reason
for this is that first-principle nonperturbative QCD calculations at finite
baryon chemical potential $\mu_B$ are not accessible by Monte Carlo
simulations, due to the presence of a complex fermion determinant in the
corresponding partition function (the so-called ``sign problem''). In this
region, which is not accessible through lattice techniques or first
principles, most of the present theoretical knowledge on the phase
transitions is obtained from the study of effective models for strong
interactions.

Given the important role played by effective models in the understanding of
the QCD phase diagram, it is important to test their reliability. This can
be done by comparing the corresponding predictions with those obtained from
first principle calculations, in situations where the latter are available.
One obvious possibility is to consider the above mentioned case of
strong-interaction matter at finite temperature and vanishing chemical
potential. Another interesting situation is the one in which $\mu_B=0$, but
one has a nonzero isospin chemical potential $\mu_I$. In this case (both at
zero and finite temperature) LQCD simulations are feasible, since the
functional determinant turns out to be real~\cite{Alford:1998sd}. Following
the early work in Refs.~\cite{Kogut:2002tm,Kogut:2002zg}, several groups
have performed LQCD calculations at $\mu_I \neq 0$ using different
techniques, see e.g.\
Refs.~\cite{Kogut:2004zg,deForcrand:2007uz,Cea:2012ev,Detmold:2012wc,Brandt:2017oyy,Brandt:2018bwq}.
One important feature confirmed by these calculations is that at $\mu_I
\gtrsim m_\pi$ one finds the onset of a Bose-Einstein pion condensation
phase, as previously conjectured in Ref.~\cite{Son:2000xc}. For a recent
review on meson condensation triggered by a large isospin imbalance see
Ref.~\cite{Mannarelli:2019hgn}, where references to various theoretical
approaches for the analysis of associated phase transitions can be found.

In this work we consider the properties of quark matter at finite isospin
chemical potential using a particular class of effective theories, viz.\ the
nonlocal Polyakov-Nambu$-$Jona-Lasinio (nlPNJL)
models~\cite{Blaschke:2007np,Contrera:2007wu,Contrera:2009hk,Contrera:2010kz,Hell:2008cc,Hell:2009by}.
In the nlPNJL approach the quarks move in a background color field and
interact through covariant nonlocal chirally symmetric four-point couplings,
which are separable in momentum space. At vanishing $\mu_B$ and finite
temperature these models provide a plausible description of chiral
restoration and deconfinement transitions, in good agreement with LQCD
results~\cite{Dumm:2021vop}. In general, it can be considered that they represent an
improvement over the local Polyakov Nambu$-$Jona-Lasinio (PNJL)
model~\cite{Meisinger:1995ih,Fukushima:2003fw,Megias:2004hj,Ratti:2005jh,
Roessner:2006xn,Mukherjee:2006hq,Sasaki:2006ww}. In fact, nonlocal
interactions arise naturally in the context of several successful approaches
to low-energy quark dynamics, and lead to a momentum dependence in quark
propagators that can be made consistent~\cite{Noguera:2008cm} with lattice
results. Moreover, it can be seen that nonlocal extensions of the NJL model
do not show some of the known inconveniences that are present in the local
theory. Well-behaved nonlocal form factors can regularize the loop integrals
in such a way that anomalies are preserved~\cite{RuizArriola:1998zi} and
charges are properly quantized. In addition, one can avoid the introduction
of various sharp cutoffs to deal with higher order loop
integrals~\cite{Blaschke:1995gr}, improving in this way the predictive power
of the models.

Within the above mentioned framework, the aim of the present work is to
provide a comparison, both at zero and finite temperature, between the
results obtained within the nlPNJL model and those arising from other
theoretical approaches. In particular, we consider the results from the
local NJL model~\cite{He:2005nk,Avancini:2019ego}, its PNJL
extension~\cite{Zhang:2006gu,Sasaki:2010jz}, chiral perturbation theory
(ChPT)~\cite{Adhikari:2019mdk} and recent LQCD
calculations~\cite{Brandt:2017oyy,Brandt:2018bwq}.

This article is organized as follows. In Sec.~\ref{model} we present the
general formalism to describe a two-flavor nonlocal PNJL model at finite
temperature and nonvanishing isospin chemical potential. In
Sec.~\ref{results} we quote and discuss our numerical results, including the
comparison with the outcomes from alternative effective approaches and LQCD
simulations. Finally, in Sec.~\ref{summary} we summarize our results and
present our main conclusions.

\section{Theoretical Formalism}
\label{model}

We start by considering the Euclidean action of a two-flavor quark model
that includes nonlocal scalar and pseudoscalar quark-antiquark currents. One
has
\begin{eqnarray}
S_E &=& \int d^4 x  \,\left[
    \bar \psi (x) \left(
    - i \rlap/\partial + \hat m  \right) \psi (x)
    \, - \, \frac{G}{2}\, j_a(x) j_a(x) \right] \ ,
\label{action}
\end{eqnarray}
where $\psi = (\psi_u\ \psi_d)^T$ stands for the $u$, $d$ quark field
doublet, and $\hat m = \mbox{diag}(m_u,m_d)$ is the current quark mass
matrix. In what follows we assume that the current masses of $u$ and $d$
quarks are equal, denoting $m_c \equiv m_u=m_d $. The nonlocal currents
$j_a(x)$ in Eq.~(\ref{action}) are given by
\begin{eqnarray}
j_a (x) &=& \int d^4 z \  {\cal G}(z) \ \bar \psi(x+\frac{z}{2}) \
\Gamma_a \ \psi(x-\frac{z}{2}) \ , \label{cuOGE}
\end{eqnarray}
where we have defined $\Gamma_a = ( \openone, i \gamma_5 \vec \tau )$,
$\tau_i$ being Pauli matrices that act on flavor space. The function ${\cal
G}(z)$ is a form factor responsible for the nonlocal character of the
four-point interactions.

To study strong-interaction matter at finite temperature and/or chemical
potential we introduce the partition function of the system, given by
$\mathcal{Z} = \int \mathcal{D} \bar{\psi}\,\mathcal{D}\psi \,\exp[-S_E]$.
As stated, we are interested in dealing with isospin asymmetric matter. This
is effectively implemented by introducing quark chemical potentials $\mu_u$
and $\mu_d$, which in principle can be different from each other. Thus, in
the effective action we perform the replacement
\begin{equation}
\left(\begin{array}{cc} \partial_4 & 0 \\ 0 & \partial_4
\end{array} \right) \rightarrow \left(\begin{array}{cc} \partial_4 -
\mu_u & 0 \\ 0 & \partial_4 - \mu_d \end{array} \right) \ .
\label{kinrep}
\end{equation}
The quark chemical potentials can be written in terms of average and isospin
chemical potentials $\mu$ and $\mu_I$ as
\begin{equation}
\mu_u \ = \ \mu + \dfrac{\mu_I}{2} \ ,\qquad\qquad \mu_d \ = \ \mu -
\dfrac{\mu_I}{2}\ ,
\end{equation}
where $\mu=\mu_B/3$, $\mu_B$ being the baryon chemical potential. For the
nonlocal model under consideration, to obtain the appropriated conserved
currents the replacement in Eq.~(\ref{kinrep}) has to be complemented with a
modification of the nonlocal currents appearing in Eq.~(\ref{cuOGE}),
namely~\cite{GomezDumm:2006vz,Dumm:2010hh}
\begin{eqnarray}
\psi(x-z/2) & \rightarrow & \mathcal{W}(x,x-z/2)  \, \psi(x-z/2)\ ,
\nonumber \\
\bar\psi(x+z/2) & \rightarrow & \bar\psi(x+z/2)\, \gamma_0
\,\mathcal{W}(x+z/2,x)\,\gamma_0 \ .
\label{transport}
\end{eqnarray}
In the present case the transport functions $\mathcal{W}$ are simply given
by
\begin{equation}
\mathcal{W}(x,x-z/2)\ = \ \mathcal{W}(x+z/2,x) \ = \ \exp \left(
\frac{z_4}{2}\, \hat \mu \right) \ ,
\end{equation}
where $\hat\mu = {\rm diag}(\mu_u,\mu_d)$.

It is convenient to perform a standard bosonization of the the fermionic
action~\cite{Ripka:1997zb}, introducing auxiliary mesonic fields $\sigma$
and $\pi_i$, $i=1,2,3$, and integrating out the fermion fields. We consider
here the mean field approximation (MFA), in which the bosonic fields are
replaced by their vacuum expectation values (VEV) $\bar\sigma$ and
$\bar\pi_i$. Let us recall that, for $\mu_I=0$, in the chiral limit
($m_c=0$) the action in Eq.~(\ref{action}) is invariant under global ${\rm
U(1)}_B \otimes {\rm SU(2)}_I \otimes {\rm SU(2)}_{IA}$ transformations. The
group U(1)$_B$ is associated to baryon number conservation, while the chiral
group SU(2)$_I \otimes {\rm SU(2)}_{IA}$ corresponds to the symmetries under
isospin and axial-isospin transformations. At zero temperature the
SU(2)$_{IA}$ symmetry is expected to be spontaneously broken by a large
value of $\bar\sigma$ (which leads to large constituent quark masses), while
at high temperatures one expects to have $\bar\sigma=0$, which implies a
restoration of the chiral symmetry. In the presence of finite quark masses
one has an explicit breakdown of SU(2)$_{IA}$ (and also of SU(2)$_I$, if
current $u$ and $d$ quark masses are different to each other), hence the
chiral symmetry is expected to be only partially restored at high $T$. Now,
in the presence of a nonzero isospin chemical potential the full chiral
symmetry group is explicitly broken down to the U(1)$_{I_3} \otimes {\rm
U(1)}_{I_3A}$ subgroup. At $T=0$ it might happen that, similarly to the
$\mu_I=0$ case, U(1)$_{I_3A}$ is spontaneously broken by a large value of
$\bar\sigma$. Moreover, while for finite current quark masses one has
$\bar \pi_3=0$~\cite{Ebert:2006uh}, it can happen that nonvanishing VEVs for
$\pi_1$ and $\pi_2$ be developed, leading to a spontaneous breakdown of the
remaining U(1)$_{I_3}$ symmetry. Since the action is still invariant under
U(1)$_{I_3}$ transformations, without loss of generality one can choose
$\bar\pi_i=\delta_{i1}\bar\Delta$.

We consider the above described general situation in which both $\bar\sigma$
and $\bar\Delta$ can be nonvanishing. At zero temperature, the mean field
thermodynamic potential is found to be given by
\begin{eqnarray}
\Omega^{\rm MFA}(T=0) &=& \frac{\bar \sigma^2 + \bar\Delta^2}{2\ G}
- {\rm Tr} \ln
\begin{pmatrix}
-\rlap/ p_u + M\big(p_u\big) & i \, \gamma_5 \, \rho\big(\bar p\big) \\
i\, \gamma_5 \, \rho\big(\bar p\big)  & -\rlap/ p_d + M\big(p_d\big)
\end{pmatrix} \ ,
\label{actionMF}
\end{eqnarray}
where
\begin{eqnarray}
M\big(p\big) \ = \ m_c + g\big(p\big)\, \bar\sigma \ ,  \qquad \qquad
\rho\big(p\big) \ = \ g\big(p\big) \, \bar\Delta \ .
\label{defsMrhop}
\end{eqnarray}
Here we have defined $p_f^\nu \equiv \left( \vec p ,\, p_4 + i \mu_f \right)$,
with $f=u,d$, and $\bar p = (p_u+p_d)/2$. The function $g(p)$ is the Fourier
transform of the form factor ${\cal G}(z)$ in Eq.~(\ref{cuOGE}).

\hfill

Let us consider the extension of the model to the case of finite
temperature, which can be addressed by using the standard Matsubara
formalism. In order to account for confinement effects, we also include the
coupling of fermions to the Polyakov loop (PL), assuming that quarks move on
a constant color background field $\phi = i g\, G_{4a} \lambda_a/2$, where
$G_{\mu a}$ are SU(3) color gauge fields. We work in the so-called Polyakov
gauge, in which the matrix $\phi$ is given a diagonal representation $\phi =
{\rm diag}(\phi_r,\phi_g,\phi_b) = \phi_3 \lambda_3 + \phi_8 \lambda_8$,
taking the traced Polyakov loop $\Phi=\frac{1}{3} {\rm Tr}\, \exp( i
\phi/T)$ as an order parameter of the confinement/deconfinement transition.
In addition, to account for effective gauge field self-interactions we
introduce a mean field Polyakov-loop potential ${\cal U}$ that depends on
the traced PL, its conjugate $\bar\Phi$ and the temperature. The resulting
scheme is usually referred to as a nonlocal Polyakov-Nambu-Jona-Lasinio
(nlPNJL)
model~\cite{Blaschke:2007np,Contrera:2007wu,Contrera:2009hk,Contrera:2010kz,Hell:2008cc,Hell:2009by}.

Concerning the PL potential, its functional form is usually based on
properties of pure gauge QCD. In this work we consider a potential given by
a polynomial function based on a Ginzburg-Landau
ansatz~\cite{Ratti:2005jh,Scavenius:2002ru}, namely
\begin{eqnarray}
\frac{{\cal{U}}_{\rm poly}(\Phi,\bar \Phi, T)}{T ^4} \ = \ -\,\frac{b_2(T)}{2}\, \bar \Phi \Phi
-\,\frac{b_3}{6}\, \left(\Phi^3 + \bar \Phi^3\right) + \,\frac{b_4}{4}\, \left( \bar \Phi \Phi \right)^2 \ ,
\label{upoly}
\end{eqnarray}
where
\begin{eqnarray}
b_2(T) = a_0 +a_1 \left(\dfrac{T_0}{T}\right) + a_2\left(\dfrac{T_0}{T}\right)^2
+ a_3\left(\dfrac{T_0}{T}\right)^3\ .
\label{pol}
\end{eqnarray}
The parameters $a_i$ and $b_i$  can be fitted to pure gauge lattice QCD
results imposing the presence of a first-order phase transition at the
reference temperature $T_0$, which is a further parameter of the model. In
the absence of dynamical quarks, $T_0$ is the critical temperature for
deconfinement, and from lattice QCD calculations one expects it to be
approximately equal to 270~MeV. However, it has been argued that in the
presence of light dynamical quarks $T_0$ should be rescaled to about 210 and
190~MeV for the case of two and three flavors, respectively, with an
uncertainty of about 30~MeV~\cite{Schaefer:2007pw,Schaefer:2009ui}. The numerical values for
the PL potential parameters are~\cite{Ratti:2005jh}
\begin{equation}
a_0 = 6.75\ ,\quad a_1 = -1.95\ ,\quad a_2 = 2.625\ ,\quad a_3 = -7.44
\ ,\quad b_3 = 0.75\ ,\quad b_4 = 7.5\ .
\end{equation}

In this way, the grand canonical thermodynamic potential of the system is
given by
\begin{eqnarray}
\Omega^{\rm MFA} & = &
\frac{\bar\sigma^2 + \bar\Delta^2}{2\ G}  - 2 \, T\ \sum_{n=-\infty}^\infty \sum_{c=r,g,b}
\int \frac{d^3 p}{(2\pi)^3} \ \ln\Big\{ E_{nuc}^2\ E_{ndc}^2 \nonumber \\
& & -\ \rho(\bar p_{nc})^2 \Big[\big(M(p_{nuc})-M(p_{ndc})\big)^2-(\mu_u-\mu_d)^2\Big] \Big\}
+ \,
{\cal{U}}_{\rm poly}(\Phi,\bar \Phi,T)\ ,
\label{granp}
\end{eqnarray}
where we have introduced the definitions $\bar p_{nc} = (p_{nuc}+p_{ndc})/2$
and $E_{nfc}^2=M(p_{nfc})^2 + p_{nfc}^2+\rho(\bar p_{nc})^2$, with $p_{nfc}
\equiv (\vec p,(2n+1)\pi T + i \mu_f +\phi_c)$. As usual in this type of
model, it is seen that $ \Omega^{\rm MFA}$ turns out to be divergent, thus
it has to be regularized. We adopt here a prescription similar as the one
considered e.g.\ in Ref.~\cite{GomezDumm:2004sr}, viz.
\begin{equation}
\Omega^{\rm MFA,\rm reg} \ = \
\Omega^{\rm MFA}\, -\, \Omega^{\rm free}_{\rm q}\,
+\, \Omega^{\rm free,reg}_{\rm q}\, +\, \Omega_0 \ .
\end{equation}
Here the ``free'' potential keeps the interaction with the PL, while $\bar
\sigma$ and $\bar \Delta$ are set to zero. A constant term $\Omega_0$ is
also added so as to fix $\Omega^{\rm MFA,\rm reg} = 0$ at $\mu_B = \mu_I = T
= 0$. For the regularized form of the free piece, the Matsubara sum can be
performed analytically. One has
\begin{eqnarray}
\Omega^{\rm free,reg}_{\rm q} = -2  T \sum_{f=u,d} \, \sum_{c=r,g,b} \,
\sum_{s=\pm 1} \int \frac{d^3 \vec{p}}{(2\pi)^3}\; \mbox{Re}\; \ln \left[ 1
+ \exp\left(-\;\frac{\epsilon_f + s\ (\mu_f + i \phi_c)}{T} \right)\right] \
,
\end{eqnarray}
where $\epsilon_f = \sqrt{\vec p^{\;2} + m_f^2}\,$.

The mean field values $\bar \sigma$ and $\bar \Delta$, as well as the values
of $\phi_3$ and $\phi_8$, can now be obtained from a set of four coupled
``gap equations'' that follow from the minimization of the regularized
thermodynamic potential, namely
\begin{equation}
\frac{\partial \Omega^{\rm MFA,reg}}{\partial \bar \sigma} \ = \ 0\
, \qquad
\frac{\partial \Omega^{\rm MFA,reg}}{\partial \bar \Delta} \ = \ 0\
, \qquad
\frac{\partial \Omega^{\rm MFA,reg}}{\partial \phi_3} \ = \ 0\
, \qquad
\frac{\partial \Omega^{\rm MFA,reg}}{\partial \phi_8} \ = \
0\ .
\label{gapeqs}
\end{equation}

In addition, it is interesting to study the behavior of quark condensates.
As usual, we consider the scalar condensate $\Sigma = \Sigma_u + \Sigma_d$,
where $\Sigma_f = \langle \bar \psi_f \psi_f \rangle$. The corresponding
expressions can be obtained by differentiating $\Omega^{\rm MFA,reg}$ with
respect to the current up and down current quark masses, i.e.
\begin{equation}
\Sigma_f \ = \ \frac{\partial \Omega^{\rm MFA,reg}}{\partial m_f} \ .
\label{Sigma}
\end{equation}
Another relevant quantity is the charged pion condensate $\Pi$, which is
expected to be nonvanishing for $\mu_I \neq 0$. According to our choice
$\bar\pi_i= \delta_{i1} \bar \Delta$, we get
\begin{eqnarray}
\Pi \ = \ \langle \bar \psi i \gamma_5 \tau_1 \psi \rangle \ .
\label{Pi}
\end{eqnarray}
The analytical expression for this condensate can be obtained by taking the
derivative of the regularized thermodynamic potential with respect to an
auxiliary parameter added to $\rho(\bar p)$ in Eq.~(\ref{actionMF}), and
then set to zero after the calculation.

To study the phase transitions, we also introduce the susceptibilities
associated to the $\Sigma$ and $\Pi$ condensates~\cite{Lu:2019diy} and the
Polyakov loop. These are given by
\begin{equation}
\chi_{\rm ch} \ = \ - \frac{\partial\Sigma}{\partial m_c} \ , \qquad
\chi_{\Pi} \ = \ \frac{\partial\Pi}{\partial m_c} \ , \qquad
\chi_\Phi \ = \ \frac{d\Phi}{dT}\ .
\label{chis}
\end{equation}

Finally, from the regularized potential one can calculate various
thermodynamic quantities, such as the energy and entropy densities
$\varepsilon$ and $s$, and the particle number densities $n_I$ and $n_B$.
The corresponding expressions are
\begin{eqnarray}
\varepsilon &=& \Omega^{\rm MFA,reg} + T\, s + n_I\, \mu_I + n_B\, \mu_B\ , \nonumber \\
s &=& -\, \frac{\partial \Omega^{\rm MFA,reg}}{\partial T} \ , \nonumber \\
n_I &=& -\, \frac{\partial \Omega^{\rm MFA,reg}}{\partial \mu_I}
\ , \nonumber \\
n_B &=& -\, \frac{\partial \Omega^{\rm MFA,reg}}{\partial \mu_B} \ .
\label{esn}
\end{eqnarray}

In this work we restrict to the case of $\mu_B = 0$, focusing on the effect
of finite isospin chemical potential $\mu_I$. As stated in the Introduction,
in this situation the results from effective models can be compared with
existing lattice QCD
calculations~\cite{Brandt:2016zdy,Brandt:2017oyy,Brandt:2017zck,Brandt:2018wkp,Brandt:2018bwq},
which do not suffer from the sign problem. Since the thermodynamic potential
turns out to be real, one gets $\Phi = \bar \Phi$, $\phi_8 = 0$, and the
last of Eqs.~(\ref{gapeqs}) is trivially satisfied.

\section{Numerical Results}
\label{results}

To fully define the model it is necessary to specify the form factor
entering the nonlocal fermion current in Eq.~(\ref{cuOGE}). In this work we
consider an exponential momentum dependence for the form factor in momentum
space,
\begin{equation}
g(p) \ = \ \exp (-p^2 / \Lambda^2)\ .
\label{ff}
\end{equation}
This form, which is widely used, guarantees a fast ultraviolet convergence
of quark loop integrals. Notice that the energy scale $\Lambda$, which acts
as an effective momentum cutoff, has to be taken as an additional parameter
of the model. Other functional forms, e.g.\ Lorentzian form factors with
integer~\cite{Dumm:2010hh} or fractional~\cite{Carlomagno:2018tyk} momentum
dependences, have also been considered in the literature.
In any case, it is seen that the form factor choice does not have
in general major impact in the qualitative predictions for the relevant
thermodynamic quantities~\cite{Carlomagno:2013ona}.

Given the form factor shape, the model parameters $m_c$, $G$ and $\Lambda$
can be fixed by requiring that the model reproduce the phenomenological
values of some selected physical quantities. If we take as inputs the pion
mass $m_\pi=138$~MeV, the pion weak decay constant $f_\pi=92.4$~MeV and the
quark condensates $\Sigma_u = \Sigma_d = - (240\ {\rm MeV})^3$, one has $m_c
= 5.67$~MeV, $\Lambda = 752$~MeV and $g = G\Lambda^2 =
20.67$~\cite{GomezDumm:2006vz}.


\subsection{Zero temperature}
\label{zeroT}

\begin{figure}[hbt]
\centering{\includegraphics[width=0.75\textwidth]{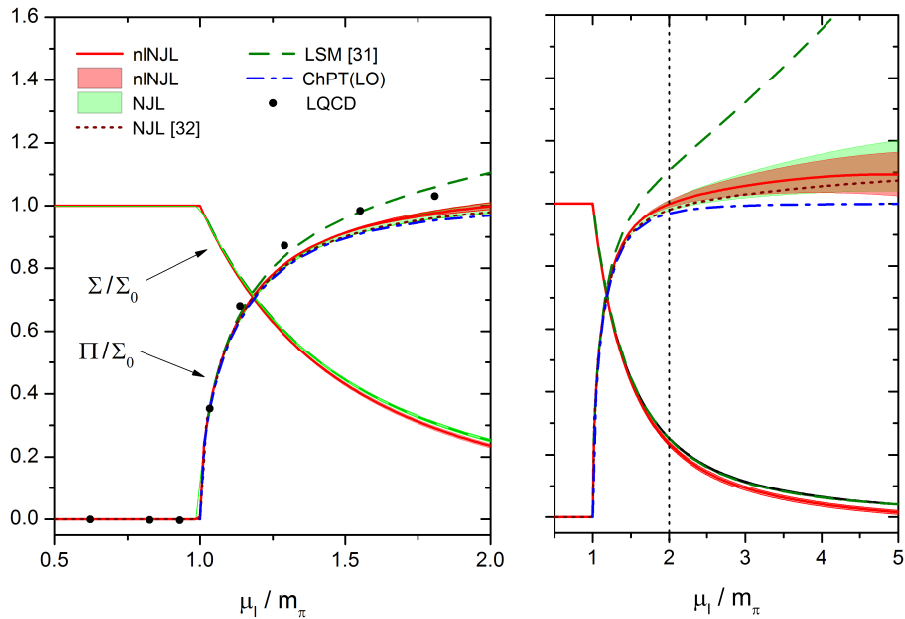}}
\caption{(Color online) Normalized $\Sigma$ and $\Pi$ condensates as
functions of the isospin chemical potential. The solid red line and the red
band correspond to the numerical results obtained within the nlNJL
model. Results from other theoretical approaches (see text) are included for
comparison.}
\label{fig:1}
\end{figure}

At zero temperature the Polyakov loop decouples from the fermions, and the
thermodynamic potential within the nonlocal NJL (nlNJL) model is given by
the expression in Eq.~(\ref{actionMF}), properly regularized. In
Fig.~\ref{fig:1} we show our numerical results for the normalized mean field
condensates $\Sigma/\Sigma_0$ and $\Pi/\Sigma_0$, where
$\Sigma_0\equiv\Sigma(\mu_I=0)$, as functions of the isospin chemical
potential. The solid red lines correspond to the parametrization described
above, which leads to $\Sigma_0 = - $(240~MeV)$^3$. To provide an estimation
of the parametrization dependence, we show with a red shaded band the
results covered by a parameter range such that $\Sigma_0$ lies between
$-$(230~MeV)$^3$ and $-$(250~MeV)$^3$. The right panel of Fig.~\ref{fig:1}
just extends the results given in the left panel, covering a broader range
of values of the scaled isospin chemical potential $\mu_I/m_\pi$. For
comparison, in both panels we include the results obtained from several
alternative approaches. The green band (partially hidden by the red one)
corresponds to the results from the local NJL, for parametrizations leading
to a quark condensate in the range between $-$(240~MeV)$^3$ and
$-$(250~MeV)$^3$. The dashed (green) lines, the dotted (brown) lines and the
dashed-dotted (blue) lines correspond to the results obtained within the
linear sigma model (LSM) in Ref.~\cite{He:2005nk}, the NJL model in
Ref.~\cite{Avancini:2019ego} (where a medium separation regularization
scheme is used) and the Chiral Perturbation Theory (ChPT) approach in
Ref.~\cite{Adhikari:2019mdk}, respectively. In addition, the fat dots denote
the results from lattice QCD obtained in Ref.~\cite{Brandt:2018bwq}.

As expected, for $\mu_I < m_\pi$ one has $\Sigma = \Sigma_0$ and $\Pi = 0$.
Indeed, for both local and nonlocal NJL models it can be analytically shown
that the onset of the pion condensation at $T=0$ occurs at $\mu_I = m_\pi$.
For larger isospin chemical potentials, as shown in Fig.~\ref{fig:1}, the
chiral condensate decreases monotonically and the charged pion condensate
gets strongly increased. In this way, for $\mu_I\geq m_\pi$ the isospin
symmetry U(1)$_{I_3}$ gets spontaneously broken, while one finds a partial
restoration of the U(1)$_{I_3A}$ symmetry for large values of $\mu_I$. From
the left panel of Fig.~\ref{fig:1} it is also seen that there is an overall
agreement between most theoretical approaches up to $\mu_I \simeq 2m_\pi$.
On the other hand, as shown in the right panel of the figure, for larger
values of $\mu_I$ there is some splitting between the predictions from
different models.

\begin{figure}[hbt]
 \centering{}\includegraphics[width=0.8\textwidth]{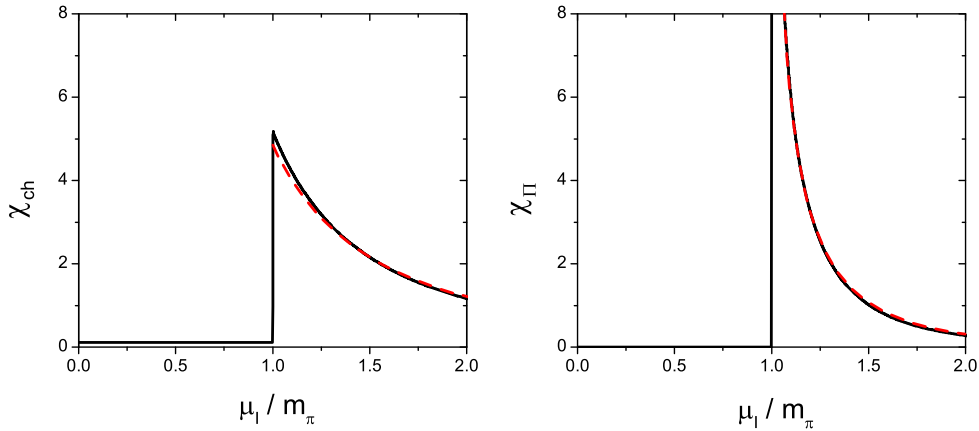}
\caption{(Color online) Chiral and pion susceptibilities as functions of the
isospin chemical potential. Solid and dashed lines correspond to the results
from nlNJL model calculations and lowest order ChPT expressions,
respectively.}
\label{fig:1b}
\end{figure}

The results for the chiral and pion condensate susceptibilities as functions
of $\mu_I$ are displayed in Fig.~\ref{fig:1b}. It can be seen that the
chiral susceptibility $\chi_{\rm ch}$ (solid line, left panel) is
approximately zero for low values of $\mu_I$, showing a jump to a high value
at $\mu_I = m_\pi$ and remaining relatively large for $\mu_I
> m_\pi$. This signals that at $\mu_I = m_\pi$ one has the onset of
a smooth transition from a phase in which the U(1)$_{I_3A}$ symmetry is
spontaneously broken to a region in which it becomes (partially) restored.
It is found that the height of the jump at $\mu_I = m_\pi$ gets increased if
the current quark mass $m_c$ is reduced. The pion condensate susceptibility
is given by the solid line in the right panel of Fig.~\ref{fig:1b}. It is
seen that $\chi_\Pi$ is zero for low values of $\mu_I$, and has a divergence
at $\mu_I = m_\pi$. This is the signature of a second order phase transition
leading to the appearance of the pion condensate, as shown in
Fig.~\ref{fig:1}. The behavior of the susceptibilities is similar to the one
found in the local NJL model, see Ref.~\cite{Lu:2019diy}

It is interesting to compare the above results with those arising from
Chiral Perturbation Theory. At the lowest order in the chiral expansion, it
is found that for $\mu_I\geq m_\pi$ the condensates satisfy the
relations~\cite{Kogut:2001id}
\begin{equation}
\frac{\Sigma}{\Sigma_0} \ = \ \frac{m_\pi^2}{\mu_I^2} \ ,
\qquad \frac{\Pi}{\Sigma_0} \ = \ \sqrt{1-\frac{m_\pi^2}{\mu_I^2}} \ .
\end{equation}
In this way one has
\begin{equation}
\left(\frac{\Sigma}{\Sigma_0}\right)^2 \, + \, \left(\frac{\Pi}{\Sigma_0}\right)^2 \ = \ 1 \ ,
\label{circle}
\end{equation}
which defines the so-called ``chiral circle''. The relation in
Eq.~(\ref{circle}) is approximately satisfied in local and nonlocal NJL
models, as can be seen from Fig.~\ref{fig:1}. In fact, the agreement is very
good up to $\mu_I \simeq 2m_\pi$, where the prediction from ChPT is
trustable. Moreover, with the aid of the Gell-Mann-Oakes-Renner relation one
can find simple analytical expressions for the susceptibilities, namely
\begin{equation}
\chi_{\rm ch} \ = \ -\,\frac{\Sigma_0}{m_c} \, \frac{m_\pi^2}{\mu_I^2}\ , \qquad
\chi_{\rm \Pi} \ = \ -\,\frac{\Sigma_0}{m_c}\,
\frac{m_\pi^4}{\mu_I^4}\, \frac{1}{\sqrt{1-\frac{m_\pi^4}{\mu_I^4}}} \ ,
\label{suscep}
\end{equation}
where it has been assumed that the ratio $\Sigma_0/f_\pi^2$ is approximately
independent of $m_c$. In Eqs.~(\ref{suscep}), it can be seen that $\chi_\Pi$
diverges at $\mu_I = m_\pi$, while $\chi_{\rm ch}$ is finite and only
becomes divergent in the chiral limit. The behavior of the susceptibilities
as functions of $\mu_I$ obtained from these equations are shown by the
dashed lines in Fig.~\ref{fig:1b}. It can be seen that they match nicely the
results arising from the nlNJL model.

\begin{figure}[hbt]
    \centering{}\includegraphics[width=0.8\textwidth]{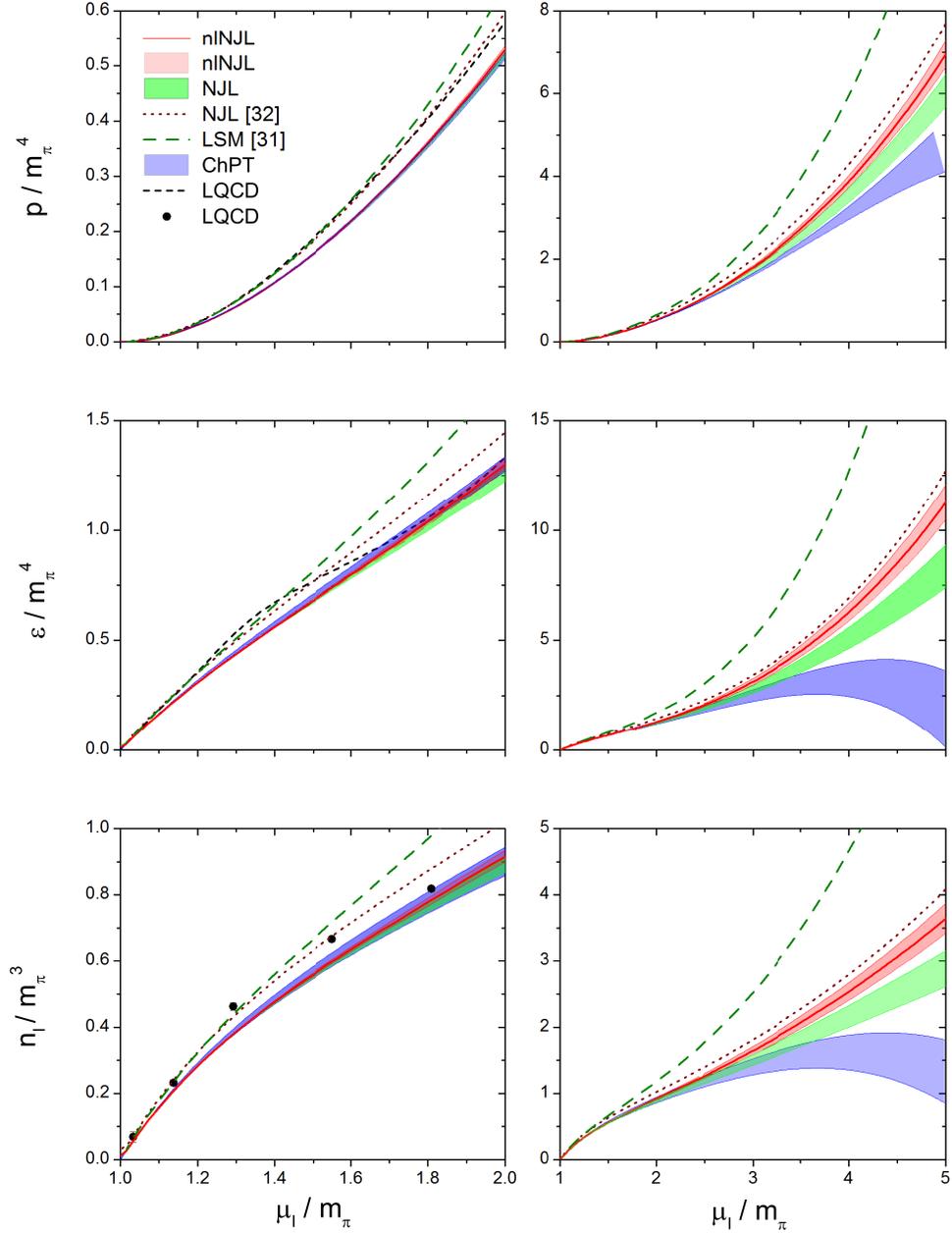}
\caption{(Color online) Numerical results for the normalized pressure,
energy density and isospin particle density as functions of the isospin
chemical potential. Besides the local and nonlocal NJL models, the graphs
include the results obtained from the ChPT approach in
Ref.~\cite{Adhikari:2019mdk}, the linear sigma model in
Ref.~\cite{He:2005nk}, and LQCD calculations in
Ref.~\cite{Brandt:2018bwq,Avancini:2019ego}.}
\label{fig:2}
\end{figure}

Next, in Fig.~\ref{fig:2} we show the results obtained within the nlNJL
model for the normalized pressure, energy density and isospin particle
density as functions of $\mu_I/m_\pi$. Results from other theoretical
approaches are also included for comparison (lines and bands for NJL and
nlNJL models are defined in the same way as in Fig.~\ref{fig:1}). In the
left panels we consider a range of $\mu_I$ from $m_\pi$ to $2m_\pi$, for
which LQCD estimations have been obtained in
Refs.~\cite{Brandt:2018bwq,Avancini:2019ego} (short-dashed black lines and
fat dots in the figure). In the right panels we include the results for the
same quantities using a different scale that covers values of the isospin
chemical potential up to $5 m_\pi$. Notice that all three quantities are
zero for $0\leq \mu_I\leq m_\pi$. From the left panels it can be seen that
in general there is a good agreement between the predictions of effective
models ---which do not differ significantly from each other--- and LQCD
results. On the other hand, for larger values of $\mu_I$ the splitting
between the results from various theoretical approaches becomes appreciably
large. Unfortunately, no LQCD results are available up to now within this
enlarged range. The behavior of the studied quantities for the nonlocal
approach (solid red lines, red bands) is found to be qualitatively similar
to the one obtained within the local NJL model (green bands), showing a
monotonic growth when $\mu_I$ gets increased. Notice that the dependence on
the parametrization turns out to be relatively low.

\begin{figure}[hbt]
\includegraphics[width=0.85\textwidth]{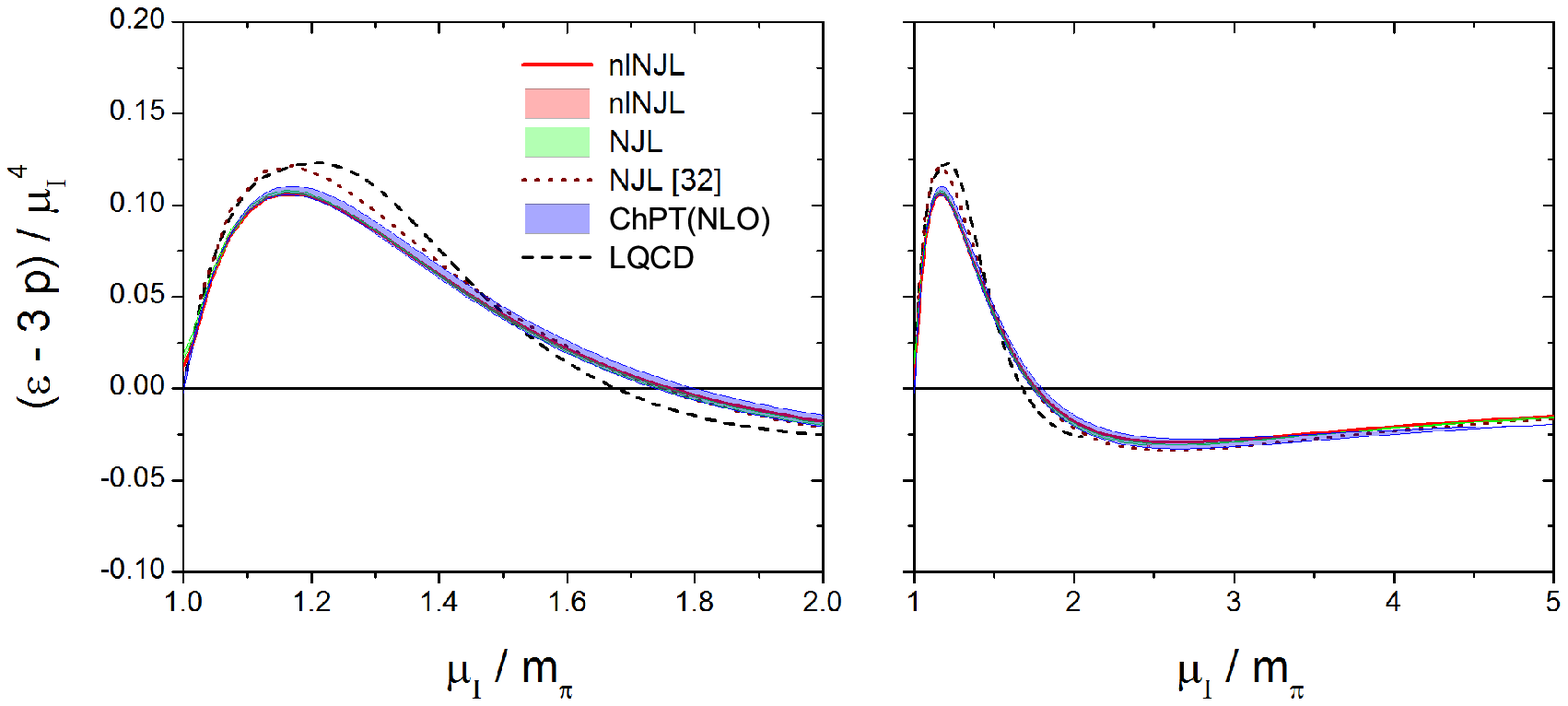}
\caption{(Color online) Numerical results for the interaction energy as
function of the isospin chemical potential. The graphs include the values
obtained from local and nonlocal NJL models, ChPT~\cite{Adhikari:2019mdk}
and LQCD calculations~\cite{Brandt:2018bwq,Avancini:2019ego}.}
\label{fig:3}
\end{figure}

Another interesting magnitude to be analyzed is the interaction energy, or
trace anomaly, $\epsilon-3p$. The behavior of this quantity (normalized by
$\mu_I^4$) as a function of $\mu_I/m_\pi$, is shown in Fig.~\ref{fig:3}. It
is seen that the results obtained within the nlNJL model are similar to
those found in other theoretical approaches. In particular, the so-called
``conformal point'', for which $\epsilon = 3p$, is reached at a value of
$\mu_I/m_\pi$ in the range between 1.75 and 1.77 (depending on the
parametrization), in good agreement with the analytical result $\mu_I/m_\pi
= \sqrt{3}$ arising from leading order ChPT~\cite{Carignano:2016rvs}.

\begin{figure}[hbt]
\vspace*{0.1cm}
\includegraphics[width=0.85\textwidth]{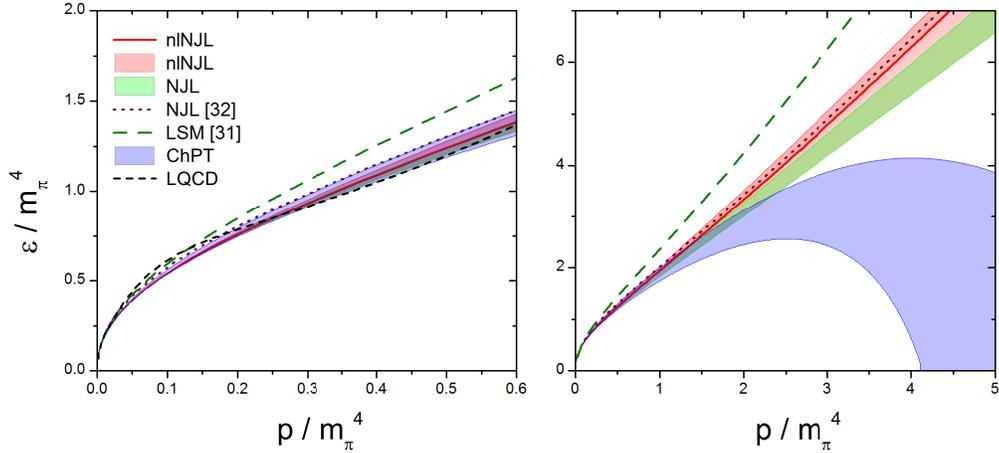}
\caption{(Color online) Numerical results for the equation of state. The
graphs include the results obtained from local and nonlocal NJL models, the
linear sigma model~\cite{He:2005nk}, ChPT~\cite{Adhikari:2019mdk} and LQCD
calculations~\cite{Brandt:2018bwq,Avancini:2019ego}.}
\label{fig:4}
\end{figure}

To conclude this subsection, in Fig.~\ref{fig:4} we plot the numerical
results obtained for the equation of state, i.e.~the behavior of the energy
density as a function of the pressure (here the isospin chemical potential
$\mu_I$ is an underlying parameter). The notation for the curves obtained
within the nonlocal NJL approach and other models, as well as those arising
from lattice QCD calculations, are the same as in Figs.~\ref{fig:2}
and~\ref{fig:3}. Once again, the results from the nonlocal approach are
qualitatively similar to those obtained in the local NJL model, and are
consistent with LQCD results in the low energy region (where LQCD data are
available).

\subsection{Finite temperature}
\label{finiteT}

We present here our numerical results at finite temperature for the
quantities defined in Sec.~\ref{model}. As discussed above, we include
the interaction between the fermions and a background color field,
considering the Polyakov loop potential in Eq.~(\ref{upoly}). The parameter
$T_0$ entering this potential is taken to be 200~MeV, following the
estimations carried out for the case of two dynamical
quarks~\cite{Schaefer:2007pw,Schaefer:2009ui}.

\begin{figure}[hbt]
   \centering{}\includegraphics[width=0.75\textwidth]{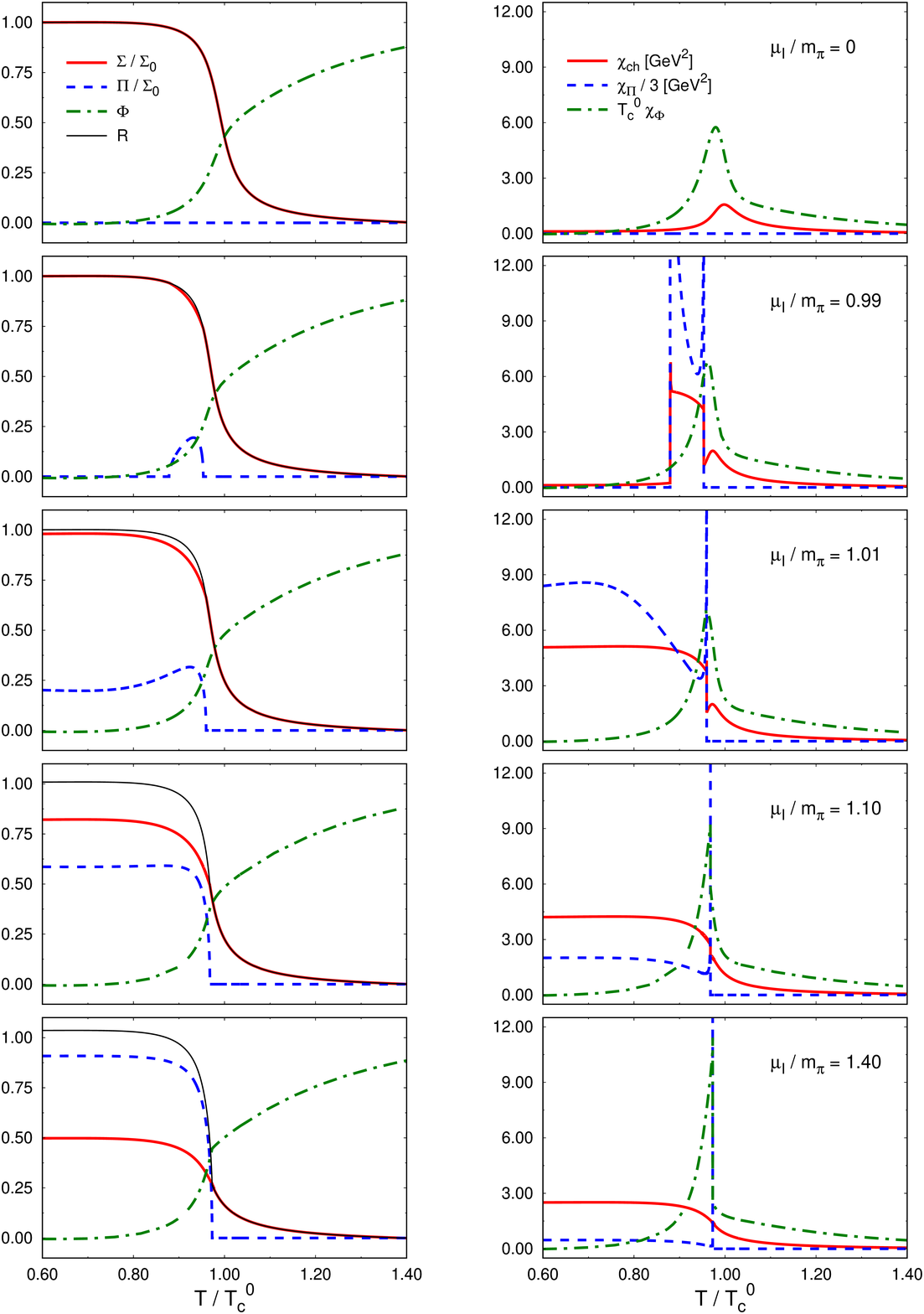}
\caption{(Color online) Left: numerical results for the normalized $\Sigma$
and $\Pi$ condensates, the traced Polyakov loop $\Phi$ and the quantity $R$
as functions of the temperature, for some fixed values of $\mu_I/m_\pi$.
Right: numerical results for the susceptibilities associated to the chiral
and pion condensates and the Polyakov loop, as functions of $T/T_c^0$.}
\label{fig:5}
\end{figure}

Let us start by studying the thermal behavior of the normalized mean field
condensates and the traced PL for some representative values of $\mu_I$
within the range $0 \leq \mu_I \leq 2m_\pi$. Our results are shown in
Fig.~\ref{fig:5}. On the left panels we plot the condensates $\Sigma$ and
$\Pi$, normalized by $\Sigma_0$ (solid and dashed lines, respectively),
together with the traced PL $\Phi$ (dashed-dotted lines). The results are
given as functions of the temperature, normalized to the critical
temperature for $\mu_I = 0$, viz.\ $T_c^0 = 174$~MeV. We also include the
curves for the normalized combined quantity $R$, defined by
\begin{equation}
R \ = \ \frac{\sqrt{\Sigma^2+\Pi^2}}{\Sigma_0}\ .
\end{equation}
In the right panels of Fig.~\ref{fig:5} we plot the susceptibilities
associated to the chiral and pion condensates and the traced Polyakov loop
(solid, dashed and dashed-dotted lines, respectively), defined in
Eq.~(\ref{chis}). As usual, the peaks of the curves for $\chi_{\rm ch}$ and
$\chi_\Phi$ are used in order to define the chiral restoration and
deconfinement transition critical temperatures.

{}From the left panels of Fig.~\ref{fig:5} it is seen that for $\mu_I=0$ the
chiral restoration and deconfinement transitions proceed as a smooth
crossovers at temperatures $T \simeq T_c^0$, while the pion condensate
vanishes for all $T$. The situation remains basically the same up to values
of $\mu_I$ approaching $m_\pi$. Then, for a small region of values of
$\mu_I$ just below $m_\pi$ (as shown explicitly for the case of
$\mu_I/m_\pi=0.99$) the pion condensate vanishes for all $T$ except for a
short range of temperatures slightly below the critical value $T_c$ that
characterizes the (almost simultaneous) chiral restoration and deconfinement
crossover transitions. On the other hand, for $\mu_I > m_\pi$, at low
temperatures the pion condensate gets nonzero values, showing the
spontaneous breakdown of isospin symmetry. These values of $\Pi$ are
approximately independent of the temperature up to $T\simeq T_c^0$, where
one finds a second order transition to a U(1)$_{I_3}$ symmetry restored
phase. In addition, it can be seen that these values of $\Pi$ get increased
with $\mu_I$, while the values of the chiral condensate $\Sigma$ decrease,
in such a way that $R$ is approximately constant. We recall that, as
discussed in the previous subsection, from lowest order ChPT one gets at
$T=0$ a constant value $R=1$ for all values of $\mu_I$. Moreover, as noted
in Ref.~\cite{He:2005nk}, the behavior of $R$ as a function of $T$ is very
similar to that found for $\Sigma/\Sigma_0$ when pion condensation is not
considered. Concerning the deconfinement transition, the graphs on the left
panel of Fig.~\ref{fig:5} show that it proceeds as a smooth crossover at an
approximately constant temperature $T \lesssim T_c^0$ for the considered
range of values of the isospin chemical potential.

\begin{figure}[hbt]
   \centering{}\includegraphics[width=0.8\textwidth]{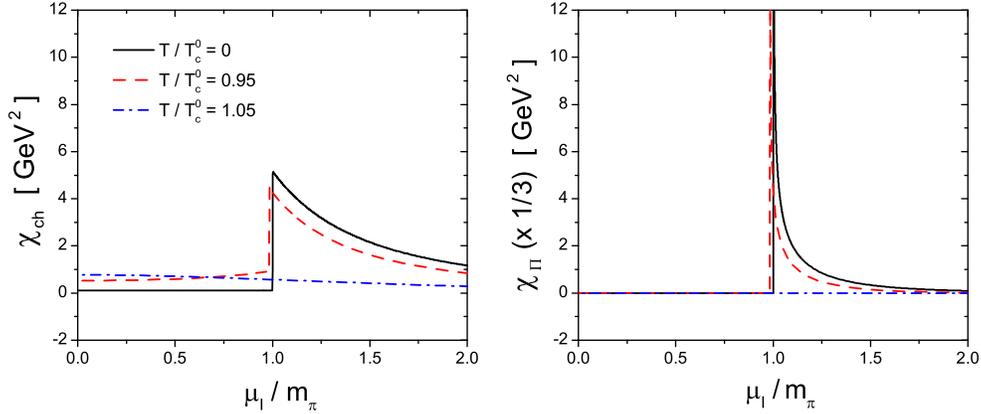}
\caption{(Color online) Chiral (left) and pion (right) condensate
susceptibilities as functions of the isospin chemical potential, for some
representative values of the temperature.}
\label{fig:5b}
\end{figure}

Taking now into account the plots in the right panels of Fig.~\ref{fig:5},
it can be seen that the PL susceptibility (green dashed-dotted lines) shows
clear peaks that indicate a crossover-like deconfinement transition at a
temperature slightly lower than $T_c^0$ and approximately independent of
$\mu_I$. In the case of the chiral susceptibility (red solid lines in the
right panels of Fig.~\ref{fig:5}), for $\mu_I=0$ one finds a peak that
defines the critical temperature $T_c^0 = 174$~MeV. Notice that for $\mu_I$
larger than $m_\pi$ the susceptibility $\chi_{\rm ch}$ is relatively large
at low temperatures. This is in agreement with the behavior shown in
Fig.~\ref{fig:1b}, and it can be attributed to the presence of a nonzero
pion condensate. The same effect occurs for values of $\mu_I$ slightly below
$m_\pi$ and temperatures $T\lesssim T_c^0$, owing to the existence of the
already mentioned nonzero value of $\Pi$ in this region (see panels of the
second row in Fig.~\ref{fig:5}). Finally, the pion condensate susceptibility
(dashed lines in the right panels of Fig.~\ref{fig:5}) is also found to be
nonzero in the presence of the pion condensate. Moreover, as expected, it
becomes divergent at the temperatures in which one finds the second order
phase transition into the isospin symmetry restored phase. These
temperatures are slightly lower than $T_c^0$ and basically coincide with the
ones corresponding to the deconfinement transition. For completeness, we
show in Fig.~\ref{fig:5b} the behavior of the $\Sigma$ and $\Pi$
susceptibilities as functions of the isospin chemical potential, for $T=0$
and temperatures slightly below and above $T_c^0$. In fact, it is seen that
the behavior of $\chi_{\rm ch}$ and $\chi_\Pi$ found for $T=0$ (see
Fig.~\ref{fig:1b}) does not change qualitatively up to the critical isospin
symmetry restoration temperature. Notice that for temperatures just below
$T_c^0$ the position of the discontinuity is shifted to $\mu_I/m_\pi$
slightly smaller than 1. It is also worth noticing that the curves for
$T\lesssim T_c^0$ are quite different from the ones obtained in the
framework of the local PNJL model, for which the discontinuity is found to
occur at significantly larger values of $\mu_I/m_\pi$ (see Fig.~3 of
Ref.~\cite{Lu:2019diy}).

\begin{figure}[hbt]
    \centering{}\includegraphics[width=0.75\textwidth]{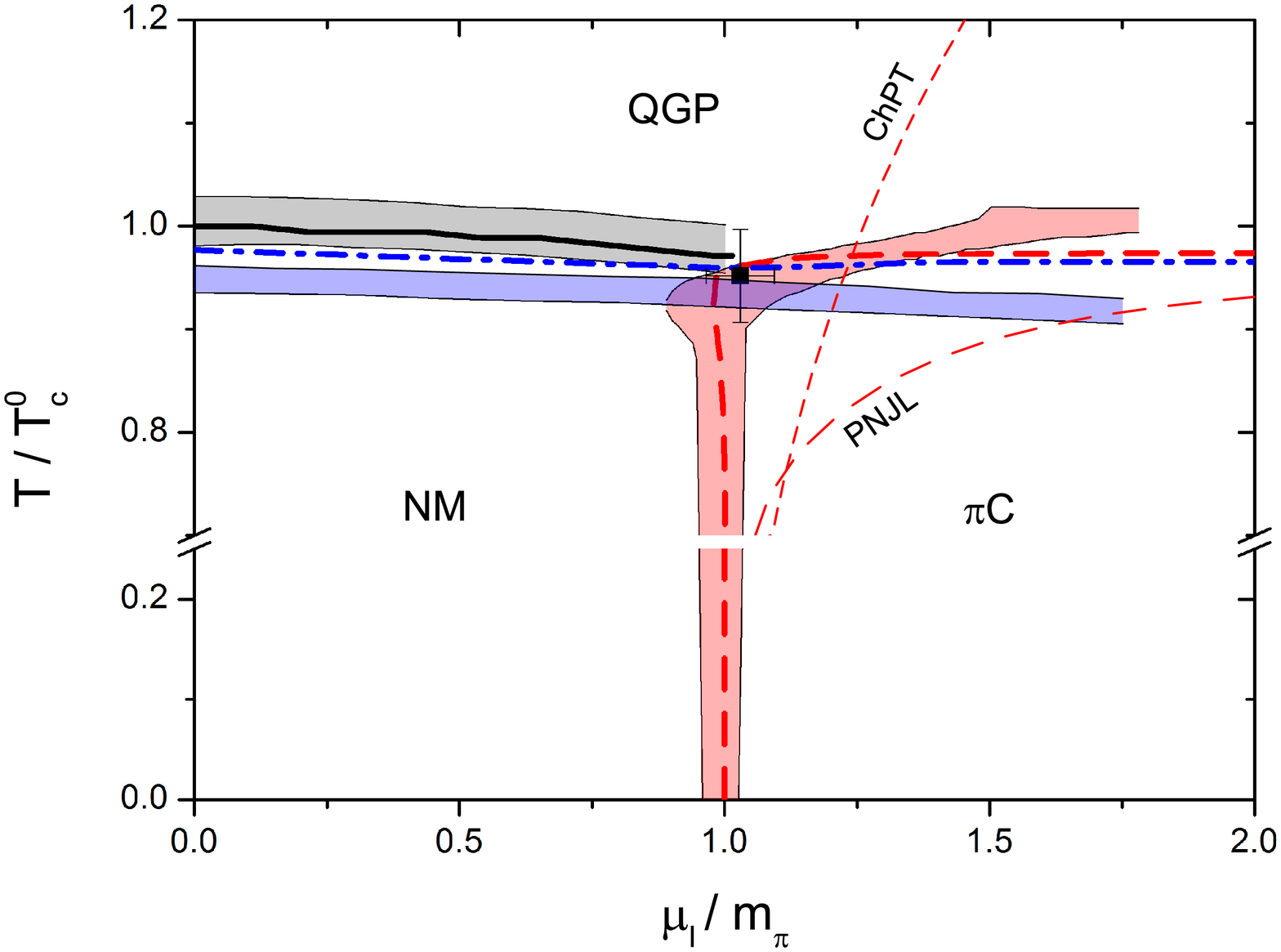}
\caption{(Color online) Phase diagram in the $\mu_I-T$ plane for the
nonlocal PNJL model. NM, QGP and $\pi$C stand for normal matter, quark gluon
plasma and pion condensation phases, respectively. Solid (black), dashed
(red) and dash-dotted (blue) lines correspond to chiral restoration, pion
condensation and deconfinement transitions, while the shaded bands indicate
the transition regions obtained from LQCD results in
Ref.~\cite{Brandt:2017oyy}. The thin dashed and short-dashed lines
indicate the NQM-$\pi$C transition curves arising from the local PNJL model
and leading order ChPT, respectively.} \label{fig:6}
\end{figure}

Through the analysis of the quantities in Fig.~\ref{fig:5} one can sketch
the phase diagram in the $\mu_I-T$ plane. This is shown in Fig.~\ref{fig:6},
where the temperature and the isospin chemical potential are normalized to
$T_c^0$ and $m_\pi$, respectively. As expected, for low values of $T$ and
$\mu_I$ the system lies in a ``normal matter'' (NM) phase, i.e.\ a
U(1)$_{I_3A}$ symmetry broken phase in which the scalar quark-antiquark
condensate $\Sigma$ is large and the pion condensate $\Pi$ is zero. By
increasing the temperature one reaches a transition to a ``quark gluon
plasma'' (QGP) phase, in which quarks deconfine and the chiral symmetry
becomes partially restored. Both chiral restoration and deconfinement
transitions occur as smooth crossovers, at approximately a common
temperature that does not depend significantly on $\mu_I$. The corresponding
curves, obtained from the peaks of $\chi_{\rm ch}$ and $\chi_\Phi$
susceptibilities, are shown by the solid and dash-dotted lines in the
figure, respectively. The results are found to be similar to those obtained
from lattice QCD calculations in Ref.~\cite{Brandt:2017oyy}, shown by the
gray and blue bands. On the other hand, for temperatures below the critical
value $T_c^0$, by increasing the isospin chemical potential one reaches a
second order phase transition to a pion-condensate ($\pi$C) region in which
the condensate $\Pi$ is nonvanishing and therefore the U(1)$_{I_3}$ symmetry
is broken. The onset of this phase, shown by the dashed line in
Fig.~\ref{fig:6}, occurs approximately at $\mu_I = m_\pi$ for all
temperature values up to $T_c^0$, in agreement with lattice QCD calculations
(red band in the figure)~\cite{Brandt:2017oyy}. Then, for $\mu_I > m_\pi$,
at a given critical temperature there is a second order phase transition
from the $\pi$C phase to the QGP phase. As discussed above, this critical
temperature is slightly lower than $T_c^0$ and remains approximately
constant for all considered values of $\mu_I$ above the pion mass. The
location of the pseudo-triple point where NM, QGP and $\pi$C phases meet is
found to be in good agreement with the result obtained in lattice QCD
calculations, given by the black square. Concerning the U(1)$_{I_3A}$
symmetry within the $\pi$C phase, it is seen that for a given temperature
$T$ lower than $T_c^0$ the values of the quark condensates decrease steadily
if $\mu_I$ gets increased beyond $m_\pi$. This can be read from the values
of $\Sigma/\Sigma_0$ shown in the left panels of Fig.~\ref{fig:5}. Notice
that for $\mu_I\simeq 1.4\,m_\pi$ the value of $\Sigma$ is found to be
reduced to approximately one half of the $\mu_I=0$ value $\Sigma_0$.
Finally, in Fig.~\ref{fig:6} we also show for comparison the $\pi$C$-$NM
transition curves corresponding to the local PNJL model and leading order
chiral perturbation theory~\cite{Splittorff:2002xn} (thin dashed and
short-dashed lines, respectively). As anticipated in the discussion
concerning Fig.~\ref{fig:5b}, it is seen that there is a substantial
difference between nonlocal and local PNJL-like approaches. This situation
does not change significantly if one considers the NJL model omitting the
interaction with the Polyakov loop.

\section{Conclusions}
\label{summary}

We have analyzed the phase diagram of strongly interacting matter
within a nonlocal two-flavor PNJL model, considering both zero and finite
temperature and nonzero isospin chemical potential. In this context, we have
studied the quark deconfinement and the breakdown/restoration of chiral and
isospin symmetries, together with the corresponding footprints on various
thermodynamic quantities.

At zero temperature, for $\mu_I = m_\pi$ one finds the onset of a phase in
which isospin symmetry is broken by the presence of a nonzero pion
condensate. Up to $\mu_I \simeq 2m_\pi$, one observes a rapid growth of this
condensate, in overall agreement with the predictions from other effective
model analyses and LQCD calculations. The agreement is also good for various
thermodynamic quantities, as the pressure, energy density, isospin particle
density and interaction energy. For larger values of $\mu_I$ (where no LQCD
data are available up to now), although one finds some general agreement in
the qualitative behavior of these quantities, there are significant
quantitative discrepancies between the results from different theoretical
approaches.

In the case of a system at finite temperature, for low values of $\mu_I$ the
pion condensate is absent and one gets, as expected, a transition from the
usual ``normal matter'' (NM) scenario into a quark-gluon plasma (QGP) phase
in which chiral symmetry is restored and quarks are deconfined. This
transition proceeds as a smooth crossover signaled by the behavior of chiral
and Polyakov loop susceptibilities. The critical temperature $T_c^0\simeq
174$~MeV is approximately the same for both chiral restoration and quark
deconfinement. For $T\leq T_c^0$, by increasing the isospin chemical
potential one finds a second order transition into a pion condensation
($\pi$C) phase, in which isospin symmetry is spontaneously broken. The
corresponding critical line $\mu_I(T)$ is a primary result of our analysis.
The critical value $\mu_I = m_\pi$ found at $T=0$ remains approximately
constant up to $T\simeq T^0_c$, reaching a pseudo-triple point in which NM,
QGP and $\pi$C phases coexist. It can be seen that there is a remarkable
agreement between these results and those obtained from lattice QCD
calculations. On the other hand, the $\pi$C-QGP transition occurs at a
temperature of the order of $T_c^0$, which is approximately constant for
$\mu_I > m_\pi$. It is worth noticing that our predictions for the border of
the $\pi$C phase region (and, in particular, for the location of the triple
point) are in good agreement with the available results from lattice QCD,
whereas they differ significantly from the predictions obtained in the
framework of the local PNJL model.

\section{Acknowledgements}

This work has been supported in part by Consejo Nacional de Investigaciones
Cient\'ificas y T\'ecnicas and Agencia Nacional de Promoci\'on Cient\'ifica
y Tecnol\'ogica (Argentina), under Grants No.~PIP17-700 and
No.~PICT17-03-0571, respectively, and by the National University of La Plata
(Argentina), Project No.~X824.



\end{document}